\documentclass{article}
\usepackage[preprint,nonatbib]{neurips_2021}

\usepackage[utf8]{inputenc} % allow utf-8 input
\usepackage[pdftex]{graphicx}

\usepackage[T1]{fontenc}    % use 8-bit T1 fonts
\usepackage{hyperref}       % hyperlinks
\usepackage{url}            % simple URL typesetting
\usepackage{booktabs}       % professional-quality tables
\usepackage{amsfonts}       % blackboard math symbols
\usepackage{nicefrac}       % compact symbols for 1/2, etc.
\usepackage{microtype}      % microtypography
\usepackage{xcolor}         % colors

\title{IMG2SMI: Translating Molecular Structure Images to Simplified Molecular-input Line-entry System}
\author{Daniel Campos \qquad Heng Ji \\ 
  Department of Computer Science \\
  University of Illinois Urbana-Champaign \\
  Urbana IL, USA \\ 
  \texttt{{dcampos3, hengji} @illinois.edu}
}
\begin{document}
\maketitle
 
\begin{abstract}
Like many scientific fields, new chemistry literature has grown at a staggering pace, with thousands of papers released every month. A large portion of chemistry literature focuses on new molecules and reactions between molecules. In this portion, the most vital information is conveyed through 2-D images of molecules, representing the underlying molecules or reactions described. In order to ensure reproducible and machine-readable molecule representations, text-based molecule descriptors like SMILES \cite{Weininger1988SMILESAC} and SELFIES \cite{Krenn2019SelfReferencingES} were created. These text-based molecule representations provide molecule generation but are unfortunately rarely present in published literature. In the absence of machine-readable molecule descriptors, the generation of molecule descriptors from the 2-D images present in the literature is necessary to understand chemistry literature at scale.  Successful methods such as Optical Structure Recognition Application (OSRA) \cite{Filippov2009OpticalSR}, and ChemSchematicResolver \cite{Beard2020ChemSchematicResolverAT} are able to extract the locations of molecules structures in chemistry papers and infer molecular descriptions and reactions. While effective, existing systems expect chemists to correct outputs, making them unsuitable for unsupervised large-scale data mining. Leveraging the task formulation of image captioning introduced by DECIMER \cite{Rajan2020DECIMERTD}, we introduce IMG2SMI, a model which leverages Deep Residual Networks \cite{He2016DeepRL} for image feature extraction and an encoder-decoder Transformer \cite{Vaswani2017AttentionIA} layers for molecule description generation. Unlike previous Neural Network-based systems, IMG2SMI builds around the task of molecule description generation, which enables IMG2SMI to outperform OSRA (Optical Structure Recognition Application) based systems by 163\% in molecule similarity prediction as measured by the molecular MACCS (Molecular ACCess System) Tanimoto Similarity. Additionally, to facilitate further research on this task, we release a new molecule prediction dataset including 81 million molecules for molecule description generation.\footnote{The programs, data and resources will be made publicly available for research purpose under a MIT License.}
\end{abstract}

\section{Introduction}
Like many other scientific research fields, chemistry has experienced an explosion of research over the last few decades. This ever-growing corpus provides ample ground for research on information extraction and text mining from vast data. Large chemistry datasets such as the patent dataset \cite{Lowe2012ExtractionOC} have been used to create models for molecule generation \cite{ArsPous2020SMILESbasedDG}, reaction yield prediction \cite{Schwaller2020PredictionOC}, and property prediction \cite{Nayak2020TransformerBM}. While these approaches have been successful, most of them have not leveraged any information learned from the corpus of chemistry papers. \\
\begin{figure}
\includegraphics[width=11cm]{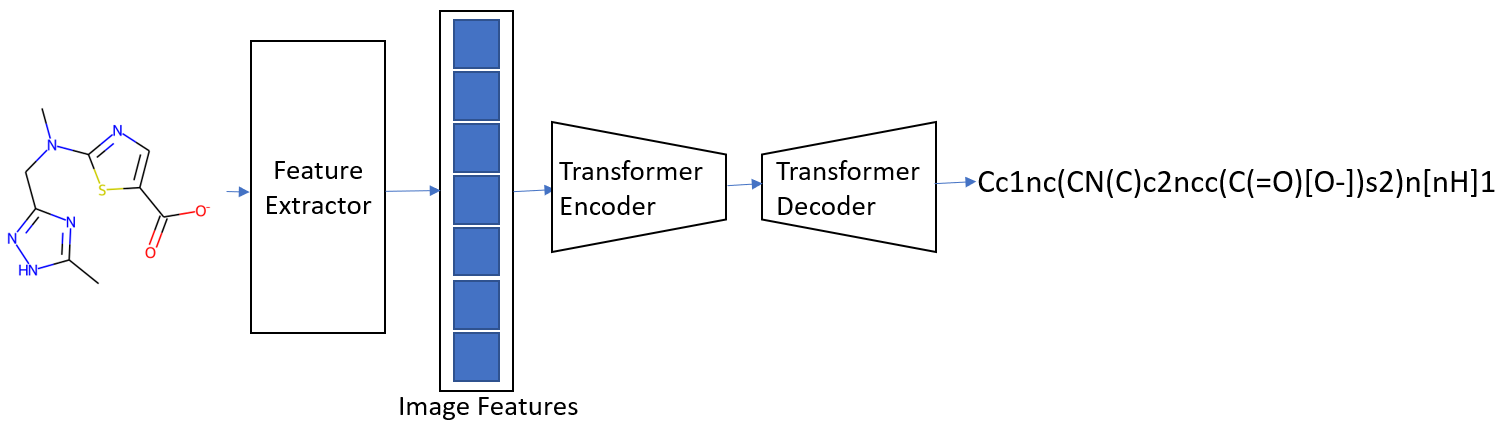}
\caption{IMG2SMI builds on popular image caption strategies leveraging a RESNET-101 based features extraction and a transformer encoder and decoder to predict a molecule description given an image.}
\label{fig:system description}
\end{figure}
While some molecules have easily recognizable names like hydrogen-peroxide ($H_2O_2$), most do not. To represent possible molecules chemists have developed a machine-readable character representation called Simplified molecular-input Line-entry System (SMILES) \cite{Weininger1988SMILESAC}. SMILES is a method of describing the structure of a molecule using ASCII strings to generate two or three-dimensional molecular representations at scale. For example,  the SMILES string "c1ccccc1" represents the molecule benzene. In recent years, deep learning researchers have focused on using large corpora of SMILES \cite{Chithrananda2020ChemBERTaLS} to learn a general representation of molecules leading to successful methods in molecule property prediction \cite{Ramsundar-et-al-2019} and molecular reaction yield \cite{Schwaller2020PredictionOC}. Despite their robust and deterministic molecule representation, SMILES are seldom found in chemistry literature as they are distinctly non-human-readable. Instead, chemistry literature focused on two-dimensional images of molecules and their reactions created by drawing programs like ChemSketch and ChemDoodle. \\
While some papers include formal molecule names or descriptions, most do not. Therefore we cannot rely only on the information embedded in the PDF. For chemists, the lack of machine-readable and reproducible molecule representations is a time sink. When chemists seek to look for related reactions or molecules in a paper, they use molecule drawing programs to recreate the molecules and then search using these representations. This approach is tedious, error-prone, time-consuming, and challenging to scale, which drives the need for an automated, high throughput molecular prediction system. Without a way to decode images, the literature is essentially full of recipes written in a language computers cannot understand. This problem is exceptionally impactful on molecule search, where reliance on manually labeled molecules and references in the text has become the standard operating methodology. Without methods of accurately extracting molecular information from images, the information conveyed in images is essentially ignored. Moreover, without accurate molecule extraction developing any method that tries to mine large corpora is likely challenging. As a result, any improvement in the quality of molecule prediction will provide amplified gains in downstream tasks.  \\
While the notion of extracting molecules from literature is by no means a novel task, most methods rely on handcrafted rules and built on the intuitions of chemists \cite{Filippov2009OpticalSR}. There are essentially two main tasks to extract visual information from chemistry literature: segmentation and molecule prediction. Segmentation systems are focused on segmenting parts of the chemistry documents and inferring what the pixel coordinates of a molecule or reaction may be. Molecule prediction systems focus on extracting the output of segmentation systems and predicting the most likely molecule for each given molecule segment found in the target document. In segmentation, a system processes a chemistry paper and produces a list of X and Y pixel coordinates representing locations where molecules and reactions are present in a paper. Description generation systems leverage the segmented images to predict a molecule description in the form of a SMILES string. As we find segmentation software to be very effective, our work focuses on molecule prediction. Despite no large dataset of annotations, molecule prediction is a perfect candidate for data-hungry solutions because artificial datasets are relatively easy to create. \\ Software tools such as RDKIT \cite{RDKIT} can use SMILES strings to create images of molecules varying style, shape, and size (pixels) at scale. While RDKIT leverages the deterministic natures of SMILES strings to create identical molecule structures each time, it can vary stylistic components such as rotation, bond highlighting, and molecule numbering to produce multiple representations of the same molecule. \\
Independently of chemistry, in the last few years computer science researchers have been able to use large data regimes and neural network architecture like transformers \cite{Vaswani2017AttentionIA} and convolutional networks \cite{Fukushima2004NeocognitronAS} \cite{LeCun1998ConvolutionalNF}, for effective image captioning \cite{Lu2018EntityawareIC} \cite{Venugopalan2017CaptioningIW} \cite{Tran2016RichIC} \cite{Anderson2018BottomUpAT} \cite{You2016ImageCW} \cite{Karpathy2015DeepVA} \cite{Xu2015ShowAA}. Image captioning has been able to leverage findings in computer vision and natural language processing to build systems which can provide descriptive, accurate, captions at an incredibly fast pace. Xu et. Al's \cite{Xu2015ShowAA} architecture of using a computer vision focused Convolutions Neural Network for feature extraction and using those features with a task-specific decoder has proven to be scalable and tune-able.
Recently, some of these approaches have begun to be applied in chemistry \cite{Rajan2020DECIMERTD}, and while these approaches have provided strong starting points, their application to chemistry is neither tuned for chemistry nor leveraging the most advanced architectures in feature extraction nor description generation. The lack of task-specific architecture modifications causes accuracy to suffer and cannot outperform traditional handcrafted systems like OSRA. \\
This paper introduces IMG2SMI, a molecule prediction approach designed for high throughput, accurate prediction of molecules. IMG2SMI is an image captioning approach that relies on a RESNET-101 \cite{He2016DeepRL} backbone for feature extraction and then a transformer \cite{Vaswani2017AttentionIA} encoder-decoder architecture for caption generation. Figure \ref{fig:system description} represents the broad IMG2SMI system architecture. While methods like DECIMER \cite{Rajan2020DECIMERTD} have been applied to molecule description generation, no previous work has modified system architecture, nor explored a non character lever representation.
Besides our model and our code, we also release a novel molecule description generation dataset called MOLCAP (\textbf{Mol}ecule \textbf{Cap}tion). MOCAP is a collection of 81 million molecules used to produce molecular images for description generation. While IMG2SMI trains with only 1 million molecules, MOLCAP is large enough and comprehensive enough for any data-hungry approach. Besides our dataset and model, in our work, we provide a quantitative and qualitative evaluation for tokenization of SMILES string using SMILES, SELFIES \cite{Krenn2019SelfReferencingES}, and Byte-Pair Encoding (BPE) \cite{Sennrich2016NeuralMT}. Our evaluation suggests that SELFIES outperforms current uses of sub-word and BPE methods that have become popular because of their success in natural language processing. Finally, as part of our work, we provide a comprehensive evaluation of evaluation methods for Molecule Description Generation finding traditional language processing metrics like ROUGE \cite{Lin2004ROUGEAP}, and Levenshtein distance combined with molecular fingerprinting provide a robust analysis method. \\
In summary, the main contributions of this paper are as follows:
\begin{enumerate}
    \item We introduce the IMG2SMI molecule prediction model. Unlike previous Neural Network-based methods, IMG2SMI is optimized for chemistry and provides a stable, accurate, and high throughput method of generating molecule descriptions for molecular images.
    \item We introduce a novel molecule description generation dataset named MOLCAP. The dataset features 81 million molecules and is large enough to enable most large data methods.
    \item We provide a thorough study of methods of processing SMILES strings for image captioning, finding SELFIES to be the most adept for image captioning.
\end{enumerate}

\section{Related Work}
\subsection{Chemistry Literature Processing}
Chemistry Literature processing is an established image processing task. Filippov et al. \cite{Filippov2009OpticalSR} create OSRA, an open-source and easy-to-use system, to understand chemistry literature and predict molecule descriptions in the form of SMILES. While effective, the OSRA system is designed for humans in the loop generation as predicted molecule descriptors require 30 seconds or less editing. This system can be useful to extract individual images of molecules for individual experimentation, it does not scale to high throughput processing. Systems like ChemReader \cite{Park2011ImagetoStructureTB}, ChemGrapher\cite{Oldenhof2020ChemGrapherOG}, and ChemSemanticResolver \cite{Beard2020ChemSchematicResolverAT} introduce improvements to OSRA by improving recognition of Bonds and R-Groups but they are still built on the traditional Optical Character Recognition structure of OSRA. To the best of our knowledge, DECIMER \cite{Rajan2020DECIMERTD} is the first system to leverage advances in computer vision and natural language processing and apply them to molecule description generation. DECIMER \cite{Rajan2020DECIMERTD} uses unaltered Show, Attend and Tell architecture \cite{Xu2015ShowAA} with a Inception V3 \cite{DBLP:journals/corr/SzegedyVISW15} based Convolutional feature extractor and a recurrent neural network-based character model for description generation. Rajan et al. prove the viability of neural-network-based methodologies applicability to chemistry literature which we leverage in our research. For further readings, please reference Rajan et Al.'s \cite{Rajan2020ARO} comprehensive review on optical, chemical structure recognition tools.
\subsubsection{Molecular Representation}
\begin{figure}[h]
\includegraphics[height=3.5cm, width=12cm]{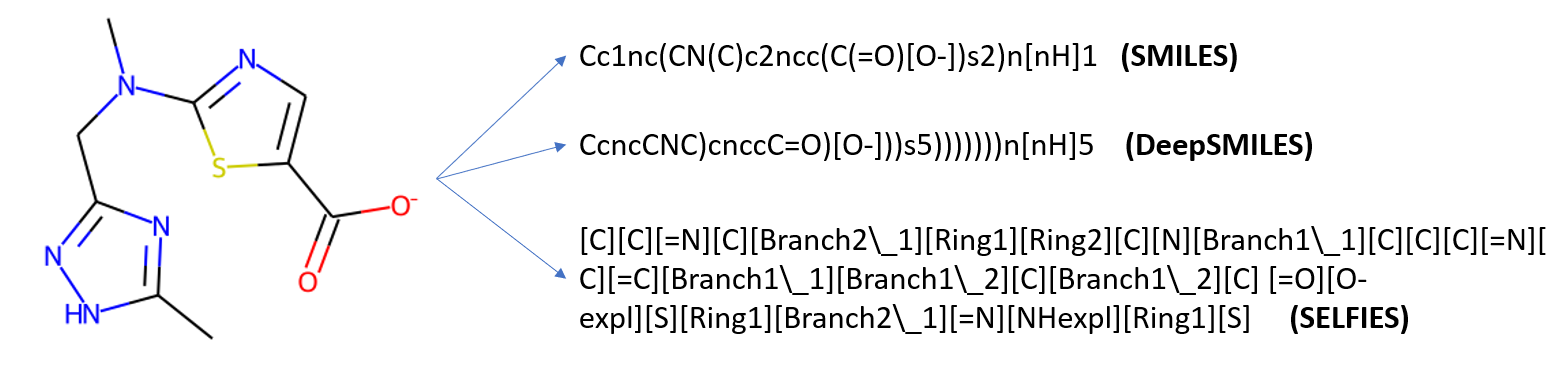}
\caption{A Molecule depicted as a two-dimensional graph where nodes are specific atoms and edges are their bonds. Each molecule has one valid SELFIES representation, one valid DeepSMILES representation, and potentially many valid SMILES representations.}
\label{fig:molecule1}
\end{figure}

The most widely used method for representing molecules in a machine-readable format is Simplified Molecular Input Line Entry System (SMILES) \cite{Weininger1988SMILESAC}. SMILES is a non-unique ASCII character-based representation used to generate molecular structure as a two-dimensional graph which is akin to how chemists would draw the molecule. SMILES are machine-readable, and most smiles strings are under 20 characters. Smiles strings are depth-first traversal of molecules graphs, and as a result, there are multiple ways to represent the same molecule. For example, C(O)C, CC(O), CCO, OCC all represent the molecular structure of ethanol. Canonicalization converts all equivalent SMILES into one representation \cite{Weininger1989SMILES2A} to simplify multiple representations for the same molecule. Additionally, not all SMILES strings are valid molecules as parenthesis must have pairs, and certain atoms do not bond. DeepSMILES \cite{OBoyle2018DeepSMILESAA} only uses closing parenthesis and introduces a single symbol for ring-closing, improving the number of possible molecule representations which map to simple molecules. SELFIES (\textbf{SELF}-referenc\textbf{I}ng \textbf{E}mbedded \textbf{S}trings) \cite{Krenn2019SelfReferencingES} is a formal Chomsky type-2 grammar with two self-referencing, recursive functions to ensure the generation of syntactically and semantically valid molecules. By flipping a single bit in a valid molecule, 100\% of SELFIES, 38.2\% of DeepSMILES, and 18.1\% of SMILES molecules are still valid. As shown in figure \ref{fig:molecule1} each molecule has associated SMILES, SELFIES, and DeepSELFIES representations which vary in length and format. As long as the string corresponds to a valid molecule, there is a deterministic, valid translation between the three molecule representations.
\subsection{Image Captioning}
Image Captioning is at the intersection of computer vision and natural language processing. While captioning systems and hybrid approaches are not a recent concept with the introduction of the COCO (Common Objects in Context) \cite{Lin2014MicrosoftCC} allowed data-intensive neural-network-based captioning approaches to thrive. Xu et al. \cite{Xu2015ShowAA} provide a successful framework leveraging an Image Classification based encoder which matched with LSTM based decoder. This Encoder-Decoder formulation has become very successful as it allows for straightforward encoder or decoder modification to understand optimal architecture. More recently, advances in Transformer's \cite{Vaswani2017AttentionIA} have been applied as decoder mechanisms such as with DETR \cite{Carion2020EndtoEndOD} or with the visual encoder as with CrossViT \cite{Chen2021CrossViTCM}. 
\subsection{Neural Networks Applied Chemistry}
The availability of large datasets like the task-specific datasets such as MoleculeNet \cite{Wu2017MoleculeNetAB}, and GuacaMol \cite{Brown2019GuacaMolBM}, and public molecule and reactions like the PubChem and ChEMBL databases has led to the broad application of neural network-based methods to chemistry. In molecule generation, methods like Generative Adversarial Networks \cite{Goodfellow2014GenerativeAN} and Variational Auto Encoders \cite{Kingma2014AutoEncodingVB} have encouraging results \cite{Brown2019GuacaMolBM} \cite{ArsPous2020SMILESbasedDG}. In molecule property prediction, Graph Convectional Networks \cite{Liu2019ChemiNetAM}, and Message Passing Neural Network \cite{Ma2020DualMP} have been used to predict quantum, physical, biophysics, physiological properties of molecules.  Other systems like ChemBERTa \cite{Chithrananda2020ChemBERTaLS} have been built on the availability of SMILES strings to create general molecule neural networks. To learn more about neural methods applied to chemistry, we suggest Elton et al.'s study on Deep Learning for molecular design \cite{Elton2019DeepLF}.
\section{Method}
\label{method}
Molecule description generation is an application of image captioning to the chemistry domain. Our model takes in an image of a molecule, X,  and generates a molecular representation of the molecule Y. The molecule image is a $N$ pixels by $M$ pixels color or image generated by a molecule drawing program such as Chemdoodle or RDKIT.  A SMILES string represents a molecule as a two-dimensional graph using a string of 1 to $C$ characters. Where c is the length of the caption and V is the length of the vocabulary. \\
\begin{equation}
    y = \{y_1, y_2..., y_c\}, y_i \in \mathbb{R}^V
\end{equation}
Building on Successful approaches in other domains, we describe the task as having separate encoding and decoding stages, each with its objective. As an image of a molecule is a two-dimensional representation, an encoder can act as a feature extractor which, given an image, produces a dense representation in a feature space. Then, leveraging the molecule representation, a decoder can produce a caption based on the extracted embedding.  
\subsection{Dataset Creation}
\begin{table}[]
\begin{tabular}{|l|l|l|l|} \hline
Dataset Name  & Size     & Original Purpose    & Average SMI Length \\ \hline
Randomized CHEMBL    & 1562045 & Molecule Generation & 47.89  \\ \hline
CHEMBL & 462301   & Molecule Generation & 54.43 \\ \hline
PubChem       & 77028926 & General Chemistry   & 44.36          \\ \hline
GDB13         & 2000000  & Molecule Generation & 22.28  \\ \hline
Decorator Scaffolds & 177024  & Molecule Generation & 39.88 \\ \hline
MOLCAP        & 81230291& Molecule Description Generation&   44.12 \\ \hline
\end{tabular}
\caption{Summary statistics of datasets used to created MOLCAP, their properties and average molecule length}
\label{tab:molcap-stats}
\end{table}
Due to lacking a large annotated dataset for experimentation, we have 
created MOLCAP. MOLCAP consists of 81 million SMILES strings mapping to 81 million molecules. To generate MOLCAP we combined existing available datasets from existing chemical databases \cite{Kim2016PubChemSA} and various experiments in molecule generation  \cite{Gupta2018GenerativeRN} \cite{ArsPous2020SMILESbasedDG} \cite{Elton2019DeepLF} \cite{ArsPous2019ExploringTG}. MOLCAP is unique as its molecules are much more complex than those which previous systems have been trained to recognize molecules under 40 characters in length. Each of the datasets used to create MOLCAP is publicly available in an open access format but licensing usage may vary for commercial applications \footnote{https://pubchemdocs.ncbi.nlm.nih.gov/downloads} \footnote{https://www.ebi.ac.uk/chembl/g/browse/compounds} \footnote{https://gdb.unibe.ch/downloads/}.\\  To create MOLCAP, first, we merge these independent datasets and keep unique molecules. Next, we convert all molecules to the canonical representation \cite{Weininger1989SMILES2A} and remove any molecules which do not have good smiles representations leaving 81,230,291 unique molecules. Table \ref{tab:molcap-stats} provides detailed statistics on the sizes and attributes of the sources we use for MOLCAP. \\
We produce character and SELFIES based dataset and following the findings of CHEMBERTA \cite{Chithrananda2020ChemBERTaLS}, we train a variety of Byte Pair Encoding (BPE) tokenizers for vocabulary size 100, 200, 500, 2000, and 20000. BPE is a hybrid between character and word-level representations, which can handle the diversity and scale of natural language corpora. BPE tokenizers are created by training custom BPE models for specific corpora size using Hugging Face's Tokenizers library \footnote{https://github.com/huggingface/tokenizers}. We produce different corpora to study how the modern tokenization approach can work with the world of molecule descriptors. BPE tokenization methods have shown to be incredibly effective in natural language as they leverage the composition nature of words and have become a popular processing method for SMILES strings. \\
Next, we select 1,000,00 molecules at random for our training corpus and 5000 for our evaluation/validation dataset, and we use RDKIT \footnote{https://github.com/rdkit/rdkit} to create a 256x256 image for each molecule and produce a tokenized caption using the six tokenization methods previously described. Our work does not study how well models deal with variations in image creation and image size as we found most molecule drawing software leverages RDKit for image creation and early experiments with augmenting data by varying size did not show major differences in output.  We have kept the evaluation portion of the dataset small as we did not see increased sensitivity with larger samples, evaluation on larger samples was slow, and this is the size of validation on other image captioning datasets such as MS-COCO. Each set of 1,000,000 molecules creates a 200GB dataset so the full image dataset for MOLCAP would be 16.24 TB which was size prohibitive for our initial experiments. 
\subsection{Evaluation Methodology}
In traditional image captioning, the evaluation metrics focus on the overlap in words present in the candidate caption compared to the reference caption. Following the standard evaluation methodology in image captioning, we employ ROUGE \cite{Lin2004ROUGEAP}, and BLEU \cite{Papineni2002BleuAM} to measure the overlap between candidates. BLEU and ROUGE measure the token overlap on tokenized SMILES strings, and as we did not see significant differences based on variation in tokenizer, we use the tokenizer with a vocabulary size of 200. Since SMILES is a character-level representation, we also explore the use of Levenshtein Distance (LD) \cite{Miller2009LevenshteinDI} as an evaluation metric. Spelling prediction or captioning task commonly uses LD to measure the edit distance in characters between the reference and caption. Additionally, we use the concept of an exact match, which we calculate by turning both target and candidate molecules into by their canonical form and searching for a direct match. It is important to note that the exact match metric is quite hard and even minor deviations in predicted molecular structure will be treated as complete failures.\\
Besides traditional Natural Language Processing methods, we evaluate models based on task-specific metrics like caption validity and caption generation. Our first metric is the image captioning percentage since in molecule description generation, systems are not always able to recognize valid molecules, and as a result, it is common for them not to generate any caption whatsoever. Our second task-specific metric is the molecular validity of generated captions. As covered in our discussion of SMILES, caption generation is complex because few SMILES strings are valid molecules. Before evaluating how good the generated molecules descriptions are, we must evaluate how often systems can generate captions and how often these captions generate valid molecular graphs.\\
Finally, we evaluate systems based on traditional molecular similarity methods using Molecule Fingerprinting and Tanimoto Similarity. Molecule fingerprinting is a common practice of representing molecule structures with bit strings using various heuristics. In our experiments, we use RDK, Morgan, and MACCS fingerprinting. RDK Fingerprinting \cite{Schneider2015GetYA} is a topological Fingerprinting is an enumeration of linear fragments from size 1 to 7, which produces a string 2048 bits long. Morgan fingerprinting \cite{Rogers2010ExtendedConnectivityF} is a form of Extended Connectivity Fingerprinting in which a molecule is decomposed into atoms and its neighbors and folded into a 2048 bit representation. MACCS fingerprinting \cite{Durant2002ReoptimizationOM} consists of representing the presence of 166 molecular substructure in each molecule. For each of these molecules, we calculate the Tanimoto Similarity as shown in equation \ref{tanimoto} (commonly known as Jaccard index), which measures the set intersection between two representations. Molecules with a Tanimoto similarity of over 0.85 have similar characteristics \cite{Bajusz2015WhyIT} and activities \cite{Dunkel2008SuperPredDC}.
\begin{equation}
    J(A,B) = \frac{A \cap B}{A \cup B} = \frac{A \cap B}{|A| + |B| -|A \cap B|}
\label{tanimoto}
\end{equation}
\section{Experiment}
\label{exp}
In order to evaluate the validity of our model and our dataset, we have trained a wide variety of models and compared their performance to the existing benchmark of OSRA \footnote{We experimented with using neural network-based methods such as DECIMER, but the model failed to generate successful captions for any molecules in the MOLCAP dataset which we attribute to the lack of task specific tuning.}. To evaluate OSRA, we leverage the python-based implementation built by Beard et al. \cite{Beard2020ChemSchematicResolverAT} using their suggested confidence level of 70 percent and captioning each image independent. In order to understand how our various metrics are related and produce a naive baseline we select a random set of 5000 molecules and compute metrics on every molecule pair in the set. This random baseline is meant to simulate the effect of choosing a molecule at random from possible molecules when generating a caption. To compare to neural methods such as DECIMER, we leveraged their open source code \footnote{https://github.com/Kohulan/DECIMER-Image-Segmentation} and their best performing pretrained model without any task specific training. 
\subsection{IMG2SMI Model Description}
As shown in Figure \ref{fig:system description} IMG2SMI consists of an image encoder, often referred to as backbone, and a caption generation decoder. Our implementation builds on Carion et al.'s \cite{Carion2020EndtoEndOD} work on image captioning. The image encoder consists of the 4th layer of a RESNET-101 \cite{He2016DeepRL} model pretrained on IMAGENET \cite{Russakovsky2015ImageNetLS}. By keeping only convolutions, the remaining model produces a 2048 dimensional vector for each molecule which represents the molecule in a dense feature space. The decoder builds on the encoder-decoder structure of Vaswani et al. \cite{Vaswani2017AttentionIA} and has three stacked layers of transformer encoders and decoders, eight attention heads, and 2048 dimensional for the feed-forward networks. We train with a batch size of 32 and train for five epochs with the dropout set to 0.1, and layer norm of 1e-12, use AdamW \cite{Loshchilov2019DecoupledWD} as our optimizer with weight decay of 1e-4, an initial learning rate of 5e-5, and the random seed of 42. Each model was trained with a single 2080ti GPU and training lasted approximately 5 hours an epoch(or about a day for a full training run). Our code builds of of the open source implementation of DETR \footnote{https://github.com/facebookresearch/detr} and model size and code configurable makes it simple to reproduce our runs and to improve on our methods. In our hyperparameter sweep we explore variations in initial learning rates of 1e-4,5e-4,1e-5,3e-5,5e-5,1-e6 finding 5e-5 to be the most optimal. We evaluated our highest performing model with a variety of random seeds including, 82, 56, 11, and 47 but saw no major variation in performance. We perform a deep ablation study where we vary the image encoder and decoder, vary the structure of the tokenized caption, and variations of the two. To vary the image encoder we experiment with fine tuning the encoder and keeping it fixed. To explore variation in the decoder we explore using LSTM + Attention identical to that which is used by Xu et al. \cite{Xu2015ShowAA}. 
\subsection{Results}
As we can see from the results in table \ref{tab:Results} IMG2SMI out performs all other existing systems outperforming the OSRA baseline MAACS Fingerprint Tanimoto Similarity (FTS) by over 163\%. When other metrics like ROUGE, a similar story is told where IMG2SMI achieves almost 10x improvement. It is worth calling out that these sizable gaps are due to the difficulty of the MOLCAP dataset. Since the median MOLCAP molecule is over 45 characters long its associated molecule is a lot larger and more complex than those which OSRA and DECIMER were tested on.\\
Moving our focus to the molecular similarity of various fingerprinting method it become quite clear how well the transformer encoder is able to create relevant molecule descriptions despite variability in input. It is worth noting that despite major changes in most metrics, Levenshtein distance stays quite high with a distance of 21. We believe this is due to IMG2SMI's tendency to produce longer captions. For example, on the evaluation set IMG2SMI's average caption is 47.66 characters compared to the reference average of 43.94. We believe with larger variation in molecular caption input and data size IMG2SMI would come closer to approximating the true average length. One surprising metric which really shows how much more work is required in the field is the exact match percentage and the BLEU score. Despite high performance in most metrics exact match is still under 10\% and BLEU score improves only by 17\%. Qualitative analysis points to minor variations in SMILES strings which have large edit distance usually caused by alterations in double bonds which produce near identical molecules with major variations in SMILES. It is worth noting that independent of what method the model uses for molecule generation (SELFIES, DeepSMILES, SMILES, BPE variations in smiles) our comparisons are performed on the translated SMILES strings. \\
It is important to compare the baseline performance to random molecule selection as this highlights the performance of existing systems. Essentially, existing methods predict molecule descriptions at a slightly better rate than choosing a molecule at random. This means that using existing systems on complex molecules is not feasible. Extending this compassion to IMG2SMI we can see that IMG2SMI is not only much better but above the target similarity of Tanimoto similarity of 0.85 where molecules tend to behave similarly. This means that unlike previous systems, on average IMG2SMI will produce a caption that has a molecule that at least behaves similar to the desired molecule. 
\begin{table}[b!]
\small
\resizebox{\textwidth}{!}{%
\begin{tabular}{|l|l|l|l|l|l|l|l|} \hline
Model                   & Exact Match(\%) & Levenshtein    & BLEU-1          & ROUGE           & MACCS FTS          & RDK FTS           & Morgan FTS  \\\hline
OSRA                    & 0.0004          & 32.76          & 0.0511          & 0.0684          & 0.3600          & 0.279  & 0.2677 \\\hline
Random Molecule         & 0.0000          & 38.32          & 0.0532          & 0.0422          & 0.3378          & 0.2229& 0.1081 \\\hline
DECIMER                 & 0.0000          & 54.00          & 0.0000          & 0.0000          & 0.0000          & 0.0000 &0.0000 \\\hline
IMG2SMI                 & \textbf{7.240}  & \textbf{21.13} & \textbf{0.0615} & \textbf{0.6240} & \textbf{0.9475} & \textbf{0.902}  & \textbf{0.8707} \\\hline
\end{tabular}}
\caption{Performance of systems for molecule description generation on MOLCAP.}
\label{tab:Results}
\end{table}
\subsubsection{Ablation Experiments}
To further study how IMG2SMI performs we explore how tokenization strategies, encoder, and decoder variation effect model performance. 
Our first experiments we vary the decoder to use a RNN + Attention as used by Xu et al. \cite{Xu2015ShowAA}. As shown in \ref{tab:ablation} the variation of decoder has a huge effect as the RNN based model is only able to slightly outperform the random molecule baseline. Building on the use of RNN or transformer we explore the effect of fixing weights for the feature extractor during training. In other words we use a feature extractor which is trained for image classification without any task specific training. While the performance of RNN + fixed encoder drop beneath our random baseline, surprisingly, the transformer and fixed encoder still outperforms all existing methods. We believe this is due to the learned shape and edge extractor providing enough of a signal to the transformers that they can learn some form of representation that comes close to approximating molecule attributes when measuring Tanimoto MACCS Similarity. The improvement found when we finetune out feature extractor provides compelling evidence that the feature extractor is able to represent molecules in a dense feature space extremely well and likely could be applied to many other tasks. \\ 
\begin{table}[b!]
\small
\resizebox{\textwidth}{!}{%
\begin{tabular}{|l|l|l|l|l|l|l|l|} \hline
Model                   & Exact Match(\%) & Levenshtein    & BLEU-1          & ROUGE           & MACCS FTS           & RDK FTS            & Morgan FTS \\\hline
Random Molecule         & 0.0000          & 38.32          & 0.0532          & 0.0422          & 0.3378          & 0.2229& 0.1081 \\\hline
IMG2SMI(RNN)            & 0.0000          & 53.02          & 0.0289          & 0.0225          & 0.1526          & 0.0954 & 0.0451 \\\hline
IMG2SMI(RNN)-F          & 0.0000          & 33.63          & 0.0549          & 0.0624          & 0.4180          & 0.2309 & 0.1328 \\\hline
IMG2SMI(Transformer)    & 2.460           & 24.70          & 0.0584          & 0.5668          & 0.7674          & 0.5724 & 0.4944 \\\hline
IMG2SMI(Transformer)-F  & \textbf{7.240}  & \textbf{21.13} & \textbf{0.0615} & \textbf{0.6240} & \textbf{0.9475} & \textbf{0.902}  & \textbf{0.8707} \\\hline
\end{tabular}}
\caption{Effects of variation of encoder and decoder on molecule description generation using SELFIES}
\label{tab:ablation}
\end{table}
To evaluate the effect of various tokenization strategies we use or tokenized datasets and train with the exact same parameters. Looking at table \ref{tab:ab2} we find tokenization method has a deep effect on molecular description generation. Few captions generated by the models using anything but selfies are actually valid molecules. Manual inspection of generated captions demonstrates similar issues as discussed by Krenn et al. \cite{Krenn2019SelfReferencingES} as many captions have unmatched parenthesis or are generally irregular bonds. Fine-tuning of the image encoder improves performance improves for all models as shown in table \ref{tab:ablation}. It is worth noting that in our experiments we attempted to alleviate this problem by leveraging beam search but this did not improve performance as increasing the beam size and branching the beam at each decoding step actually led to decreased performance. Using these results we recommend future researchers focus their efforts on SELFIES as it is likely a more fruitful representation.  If researchers must use some for of BPE tokenization we recommend a vocabulary size of 2000 as it on average produces the best results.
\begin{table}[]
\small
\begin{tabular}{|l|l|l|l|l|l|l|}
    \hline
        Vocab & Fine-tune Encoder & Images Captioned (\%) & Valid Captions (\%) \\ \hline
        OSRA(Baseline) &N/A & 86.5 & 65.2 \\ \hline
        SELFIES & No& 61.8 & 61.1  \\ \hline
        SELFIES & Yes & \textbf{99.4} & \textbf{99.4} \\ \hline
        Character & No & 1.3 & 1.0 \\ \hline
        Character & Yes & 2.2 & 2.1 \\ \hline
        BPE-100 & Yes&  3.7 & 3.5 \\ \hline
        BPE-100 & No& 2.9 & 2.9 \\ \hline
        BPE-200 & No&  3.7 & 3.5  \\ \hline
        BPE-200 & Yes& 21.5 & 21.1   \\ \hline
        BPE-500 & No&  3.3 & 3.1  \\ \hline
        BPE-500 & Yes& 8.7 & 8.5  \\ \hline
        BPE-2000 & No& 10.2 & 10.0  \\ \hline
        BPE-2000 &Yes& 20.2 & 20.0 \\ \hline
        BPE-20000 &No& 5.5 & 5.4 \\ \hline
        BPE-20000 &Yes& 18.1 & 18.0  \\ \hline
    \end{tabular}
    \caption{Effects of SMILES Tokenization Strategy. SELFIES vastly outperforms other systems but if a BPE is to be used a vocabulary size of 2000 performs best when average with fixed and fine-tuned encoder}
    \label{tab:ab2}
\end{table}
\subsection{Metrics for Molecule Description Generation}
Broadly focusing on the variation found in our evaluation metrics, we can see that some metrics are more suited for evaluation than others. Despite massive variations in other metrics, BLEU performance is virtually unchanged across systems and a poor predictor of molecular similarity. Unlike BLEU, ROUGE provides a more sensitive metric showing nearly a 10x difference between OSRA and IMG2SMI. We believe the difference between BLEU and ROUGE is related to the low exact match shared across systems. IMG2SMI recalls almost 63\% of tokens but has a low precision indicating that IMG2SMI is including additional tokens. Molecule fingerprinting metric agree in direction but diverge in magnitude which we attribute to the varying sensitivity of each method. MACCS only had 166 (bits) substructures thus is a simpler representation compared to Morgan or Path(2048 bits). It is worth noting fingerprinting methods metrics are recall focused as they measure the shared presence of a substructure and does not account for potential duplication's of a structure. 
\subsection{Limitations}
\label{limitations}
Despite competitive accuracy on molecular similarity metrics, even the best performing IMG2SMI model shows a sizable gap in terms of exact match, ROUGE, and LD. As IMG2SMI cannot have a complete exact match on even 10\% of images we believe model performance is likely able to improve drastically. It is also important to note that unlike traditional image processing methods, neural-network based methods do not perform well on data that falls outside of the training distribution. Since MOLCAP is built around complex molecules and the average molecular length > 40 tokens, the model does not provide good captions for short molecules. Since traditional systems like OSRA do well with short molecules we believe using the two systems in conjunction is likely to yield the best results. In the future we seek to improve IMG2SMI and MOLCAP by providing a more broad distribution of molecule sizes.
\section{Conclusion and Future Work}
Our experiments have shown that IMG2SMI leverages advances in Computer Vision, Chemistry AI, and Natural Language Processing to produce a molecule descriptor generation system that is designed for unsupervised, high throughput usage. When paired with existing segmentation methods, IMG2SMI can predict molecule descriptions with an accuracy unseen until now. It is important to note that while our system outperforms existing non-neural methods by 163\% and the Tanimoto similarity is 0.9475, the amount of exact matches extract is 7.24\%. The broad gap between Tanimoto similarity and exact match indicates how much better systems can become. Given the extensible nature of the architecture of IMG2SMI, we believe that our method provides a robust baseline for future research. With our release of the MOLCAP dataset, we provide the Document Analysis community with a relevant and competitive benchmark which will hopefully catalyze increased interest in chemical document analysis. This dataset is large enough to enable the most data-hungry methods in pre-training, transfer learning, and image augmentations. Finally, our exploration in molecule representations is likely to help other researchers find optimal ways of exploring the chemical space. Our findings provide empirical proof that the SMILES methods can prove difficult to use with diverse data regardless of tokenization methods, and SELFIES-based methods perform better.   \\
In the future, we plan to continue our work to address current shortcomings. We will expand the training corpus from the 1 million we used to the 81 million in MOLCAP and employ image augmentation methods such as cropping, rotation, and noise injection to build a more robust model. Moreover, we will combine IMG2SMI with existing document segmentation models to produce an easy-to-use end-to-end chemistry extraction tool. We plan to integrate this tool into a web-based extension to allow chemistry researchers to extract molecule names when reading papers automatically. Additionally, inspired by the ability to extract relevant chemistry information using the RESNET-101 backbone, we plan to include the visual features in other chemistry tasks. We believe that leveraging the extracted visual features of molecules, we will be able to use these features in other molecule property prediction tasks such as yield prediction, property prediction, and molecule substitution as most methods are not multi modal and rely only on SMILES based representations  \cite{Nayak2020TransformerBM}.
\section*{Broader Impact}
\label{broader_impact}
Our work has the potential impact to enable large scale mining of chemistry literature. By producing a accurate, high throughput experimentation method we provide the broader community with an effective way of mining existing literature at scale. IMG2SMI will allow for exploration in what kind of information can be extracted in an unsupervised way from decades of chemistry literature but since exact match accuracy is low there will be wrong predictions for many molecules
\bibliographystyle{splncs04.bst}
\bibliography{sources}
\end{document}